\documentclass[aps,prb,twocolumn,floats]{revtex4}
\usepackage{color,graphicx,pstricks}
\begin{document}  

\title {Interfacial Magnetism in Manganite Superlattices}

\author{Kalpataru Pradhan and Arno P. Kampf}

\affiliation{Center for Electronic Correlations and Magnetism, 
Theoretical Physics III, Institute of Physics, University of Augsburg, 
D-86135 Augsburg, Germany}

\date{\today}

\begin{abstract}
We use a two-orbital double-exchange model including Jahn-Teller lattice distortions, superexchange 
interactions, and long-range Coulomb (LRC) interactions to investigate the origin of 
magnetically disordered interfaces between ferromagnetic metallic (FM) and antiferromagnetic 
insulating (AFI) manganites in FM/AFI superlattices. The induced magnetic moment in the AFI 
layer varies non-monotonically with increasing AFI layer width as seen in the experiment. 
We provide a framework for understanding this non-monotonic behavior which has a one-to-one 
correspondence with the magnetization of the FM interface. The obtained insights provide a basis
for improving the tunneling magnetoresistance in FM/AFI manganite superlattices by
avoiding a magnetic dead layer (MDL) in the FM manganite.
\end{abstract}

\maketitle

The FM manganites have emerged as potential candidates for spintronics 
devices\cite{teresa-manganite-spintronics,bibes-mtj-spintronics} due to their high spin 
polarization\cite{park-lsmo-halfmetal,tok-book}. For the future generation of magnetic tunnel 
junctions (MTJs) artificial trilayers of insulating metal oxides sandwiched between FM 
manganites are currently designed. In MTJs a large tunneling magnetoresistance (TMR)\cite{bowen-tmr-large} 
is observed by switching the spin orientation in the FM leads from antiparallel to parallel 
configurations; the TMR is defined by the ratio $(R_{AP}-R_P)/R_{P}$ where $R_{AP}$ and $R_{P}$ 
are the resistances for antiparallel and parallel orientations, respectively\cite{julliere-tmr}. 
Although SrTiO$_3$ is predominantly used as the insulator between La$_{0.67}$Sr$_{0.33}$MnO$_3$ (LSMO) 
layers, also other combinations of FM and NMI oxides 
(FM = LSMO, La$_{0.67}$Ca$_{0.33}$MnO$_3$ (LCMO); NMI = TiO$_2$, LaAlO$_3$, NdGaO$_3$) have 
also been tested for their performance\cite{sun-lcmo-tmr,sun-NdGaO3-mdl,bibes-lcmo-mdl,bibes-tio2-tmr}. 

TMR is a spin dependent process which critically depends on the magnetic and the electronic properties 
of the interface between FM manganites and the insulating material\cite{teresa-manganite-spintronics}. 
In such a spin sensitive device it is required to have a structurally and magnetically well defined 
interface. It is an experimental fact that the magnetization of FM manganites decreases at the interface 
below its bulk value\cite{freeland-magnetization-interface}, the origin of which is not well understood. 
The reduction of the magnetization at the interface, usually referred to as the `magnetic dead layer' 
(MDL)\cite{sun-NdGaO3-mdl,bibes-lcmo-mdl} has an adverse effect on the TMR by decreasing the tunneling 
current, which by itself should be large for device applications. A recent microscopic
analysis\cite{mdl-lsmo-sto} suggests that the decrease of the double-exchange energy at an FM/SrTiO$_3$
interface is the origin of the MDL. The possible coexistence of different magnetic phases at the interface
is however not accessible from such an analysis.

The reduction of the magnetization has been attributed to phase separation, and/or electronic and magnetic 
reconstructions due to structural inhomogeneities at the interface. Unlike at the surface of FM 
manganites\cite{park-lsmo-halfmetal} it is a difficult task to determine the electronic and structural 
changes at interfaces which are several nanometers below the surface. To minimize disorder and strain effects 
isostructural interfaces are favorable. In a different approach NMI barriers were 
replaced by AFI manganites\cite{cheng-tri-layer,li-lcmo-pcmo,jo-lcmo-lcmo,mathur-lsmo-pcmo}. 
In the presence of a small external magnetic field not only the FM manganites align but there is a likely 
possibility that the magnetization in the AFI layer also aligns along the FM leads\cite{li-lcmo-pcmo,mathur-lsmo-pcmo}. 
Specifically the relation between the magnetoresistance and the induced magnetic moment in the AFI barrier 
was established in LSMO/Pr$_{0.67}$Ca$_{0.33}$MnO$_3$(PCMO)/LSMO superlattices\cite{mathur-lsmo-pcmo1}. 
The magnetic moment of the PCMO layers in the superlattice behaves non-monotonically with increasing PCMO 
layer width\cite{mathur-lsmo-pcmo}. Remarkably the magnetoresistance follows a very similar non-monotonic 
behavior. It is a priori not clear from the LSMO/PCMO/LSMO superlattices, if MDLs at the interface exist 
for different widths of the PCMO layers. 

In this letter, we explore in detail the electronic and magnetic reconstructions of the FM/AFI 
superlattices at the electron density $n = 0.5$ for different widths of the AFI layers. Electrons are 
transferred from the FM to the AFI layers at the interface even though the initial electron density in 
the bulk materials are equal. The amount of electron transfer from the FM interfacial line 
depends upon the thickness of the AFI layer. We explain the non-monotonic behavior of the induced 
ferromagnetic moment in the AFI layer with increasing AFI layer width and establish explicitly a 
one-to-one correspondence between the induced magnetic moment in the AFI layer and the magnetization 
at the interface in the FM/AFI superlattices. This concept establishes a route to minimize or
even avoid the MDL in FM/AFI superlattices.

We consider a two-dimensional model Hamiltonian for manganite superlattices composed of alternating 
FM and AFI regions. The model and the method we employ have been elaborately discussed in 
Ref.\citenum{kp-sl-prb}. The model is given by 
\begin{eqnarray}
H = H_{FM} + H_{AFI} + H_{lrc}, 
\end{eqnarray}
where both $H_{FM}$ and $H_{AFI}$ have the same reference Hamiltonian
\cite{yunoki-hamil,dag-hamil-pr,sanjeev-x-0,kp-bsite-prl}
\begin{eqnarray}
H_{ref} &=& \sum_{\langle ij \rangle \sigma}^{\alpha \beta}
t_{\alpha \beta}^{ij}
 c^{\dagger}_{i \alpha \sigma} c^{~}_{j \beta \sigma}
 - J_H\sum_i {\bf S}_i\cdot{\mbox {\boldmath $\sigma$}}_i
+ J\sum_{\langle ij \rangle} {\bf S}_i\cdot{\bf S}_j \cr 
&&- \lambda \sum_i {\bf Q}_i\cdot{\mbox {\boldmath $\tau$}}_i
+ {K \over 2} \sum_i {\bf Q}_i^2 - 
{\mu\sum_{i \alpha \sigma} c^{\dagger}_{i \alpha \sigma} c^{~}_{i \alpha \sigma}}. 
\end{eqnarray}
\noindent

$H_{ref}$ is constructed to qualitatively reproduce the phase diagram in the bulk 
limit\cite{kp-bsite-prl,kp-bsite-epl}. $\lambda$ is the coupling between the e$_g$ electron and the 
Jahn-Teller phonons ${Q}_{i}$ in the adiabatic limit, and $J$ is the superexchange interactions between 
the $t_{2g}$ spins ${\bf S}_i$. We treat ${\bf S}_i$ and ${Q}_{i}$ as classical\cite{class-ref1} and 
set $|{\bf S}_i|=1$. We also set the stiffness of Jahn-Teller modes $K=1$ and the Mn-Mn hopping $t=1$. 
$t$ is the reference energy scale. 
 
The Hund$'$s coupling $J_H$, between ${\bf S}_i$ and the $e_g$ electron spin {\boldmath $\sigma$}$_i$, 
estimated to be 2 eV\cite{okimoto-hunds} in manganites, is much larger than 
$t$ ($t\sim0.2-0.5 eV$\cite{satpathy-t}). For this reason we use the limit 
$J_H \rightarrow \infty$\cite{dag-hamil-pr}. In an external magnetic field $h$ we add a Zeeman coupling 
term $H_{mag} = -{\bf h}\cdot\sum_i {\bf S}_i$ to the Hamiltonian 

The average electron density of the FM/AFI superlattice is fixed by choosing the same chemical potential 
$\mu$ at each site. The LRC part $H_{lrc} = \sum_{i} \phi_i n_i$ of the Hamiltonian 
controls the amount of charge transfer across the interface. Here a self-consistent solution of the 
Coulomb potentials 
${\phi}_{i} = \alpha t \sum_{j\neq i} \frac{\langle n_{j} \rangle-Z_j}{|{\bf R}_{i}-{\bf R}_{j}|}$ 
is set up at the mean-field level\cite{millis-mft-int,yunoki-poisson-int,brey-0.67-0.67}; for details 
see Ref.\citenum{kp-sl-prb}. $\alpha$ = $e^2$/$\epsilon$$at$ is the Coulomb interaction strength 
where $\epsilon$ and $a$ are the dielectric constant and the lattice parameter, respectively. 
For the 2D case considered here $\alpha$ is approximately 0.1\cite{kp-sl-prb}.

We use the `traveling cluster approximation' (TCA)\cite{tca-ref} based Monte Carlo sampling technique. This 
method was already successfully applied in several earlier studies \cite{sanjeev-x-0,kp-bsite-prl,kp-bsite-epl}. 
At each system sweep, with the additional $H_{lrc}$ term in the Hamiltonian, we solve for the Coulomb potentials 
${\phi}_{i}$ self-consistently until the electron density $n_i$ at each site is converged\cite{kp-sl-prb}. 
All physical quantities are averaged over the results for ten different `samples' where each sample denotes 
a different initial realization of the classical variables. 

Here we analyze specifically superlattices composed of FM and AFI manganites of equal electron density 
$n=0.5$. We use the typical value $J=0.1$\cite{perring-se,kp-bsite-prl} for both the FM and the AFI 
manganites and differentiate between a FM and an AFI phase by varying $\lambda$. For the parameters 
$J=0.1$ and $n=0.5$, the groundstate is a FM for $\lambda \equiv \lambda_M = 1.0$ while it is an AFI for 
$\lambda \equiv \lambda_I \geq 1.6$. The AFI phase at $n = 0.5$ is a charge and orbital ordered CE 
phase\cite{kp-bsite-prl}. The density of states is finite for the FM phase while it is gapped at the 
Fermi level for the AFI phase; charge transfer from the FM to the AFI side is expected when the FM and 
the AFI are joined together.

Two types of FM/AFI superlattices are shown schematically in Fig.1; $w$ denotes the width of the AFI spacer. 
Periodic boundary conditions in both directions ensures that the superlattice structures are composed of 
alternating FM and AFI layers. The type I superlattice is considered in the following discussions while 
results for the type II superlattice are discussed in the concluding paragraphs. 

\begin{figure} [t]
\centerline{
\includegraphics[width=8.75cm,height=4.25cm,clip=true]{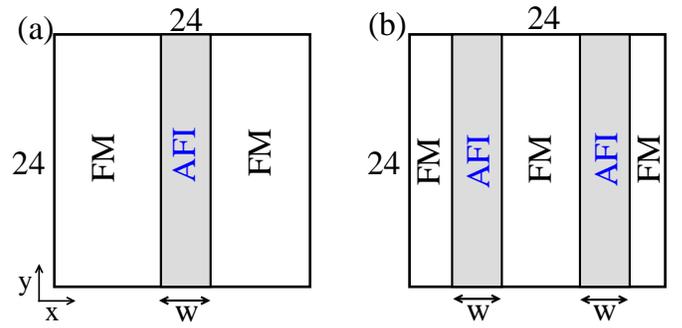}}
\caption{(Color online) Schematic view of the FM/AFI superlattice on a 24$\times$24 
lattice. (a) Type I : one AFI layer (b) type II: two AFI layers. 
}
\end{figure}

Different combinations of electron-phonon couplings ($\lambda_M=1.0$, $\lambda_I=1.6-2.0$) are 
considered first in the absence of LRC intreractions $\alpha=0$. To start with we discuss the results 
for $\lambda_I=$ 1.65 (see Fig.2(a)), for which the ferromagnetic structure factor 
$\langle S_I(\textbf{0}) \rangle$ behaves non-monotonically with increasing AFI layer width, 
where $S_I(\textbf{q})$ =${1 \over N_I^2}$ $\sum_{ij \in AFI}$ 
$\bf {\bf S}_i\cdot {\bf S}_j$ e$^{i\bf{q} \cdot ({\bf r}_i-{\bf r}_j)}$ and the angular bracket denotes 
the average over thermal equilibrium configurations combined with an additional average over ten different 
`samples'. The induced magnetization in the AFI layer is small for $w=1$, nearly equal 
to 1 (all the t$_{2g}$ spins are fully ordered) for $w=2$, and rapidly decreases for $w>7$. 

The averaged $z$ component of the t$_{2g}$ spins $\langle S_{zI} \rangle$ in the AFI layer for 
$\lambda_I$ = 1.65 is similarly non-monotonic as $\langle S_I(\textbf{0}) \rangle$ as shown in 
Fig.2(b). We also calculate the local staggered charge order by 
$\langle CO_I \rangle $ = ${1 \over N_I}$ $\sum_{i \in AFI}$ $\langle n_i \rangle$e$^{i(\pi,\pi) 
\cdot {{\bf r}_i}}$ where $i$ denotes lattice sites in the AFI layer with position ${{\bf r}_i}$. 
$\langle CO_I \rangle$, shown in Fig.2(b), remains small for $w\leq7$ and starts to rise for $w>7$. 
The decrease in the magnetization accompanied by the emerging charge order indicate that the AFI layer 
gradually returns to the bulk AFI state with increasing $w$.

\begin{figure} [t]
\centerline{
\includegraphics[width=8.75cm,height=7.75cm,clip=true]{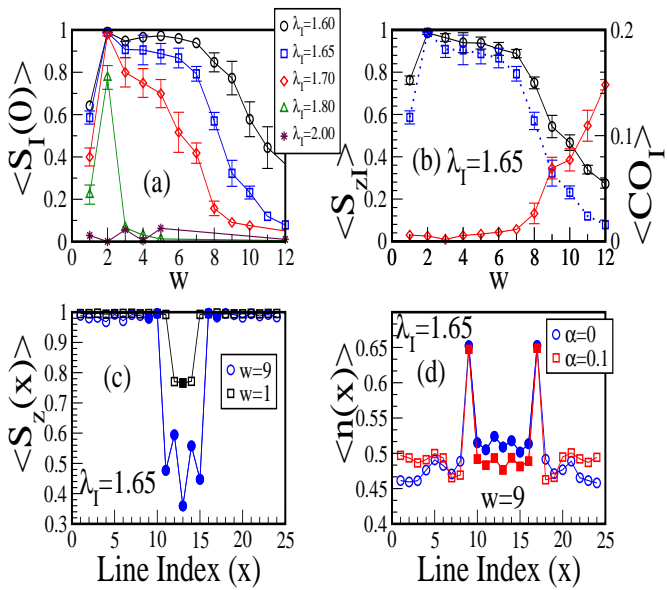}}
\caption{(Color online) (a) Ferromagnetic structure factor  $\langle S_I({\textbf 0}) \rangle$ in the 
AFI layer for $w=1-12$ at $T=0.01$ ($\lambda_M=1.0$ and $\lambda_I=1.6-2.0$). 
(b) $\langle S_{zI} \rangle$ and $\langle CO_I \rangle$ (see text) in the AFI layer for $\lambda_M=1.0$ 
and $\lambda_I=1.65$. $\langle S_I({\textbf 0}) \rangle$ is also included as the dotted line.
(c) Line averaged $z$ component of the t$_{2g}$ spins $\langle S_z(x) \rangle$ for $w=9$ and $w=1$, 
(d) line averaged electron density $\langle n(x) \rangle $ for $w=9$ with ($\alpha=0.1$) and 
without ($\alpha=0.0$) LRC interactions. In (c) and (d) open and closed symbols are from lines 
in the FM and the AFI layers, respectively. 
}
\end{figure}

\begin{figure} [t]
\centerline{
\includegraphics[width=8.75cm,height=7.75cm,clip=true]{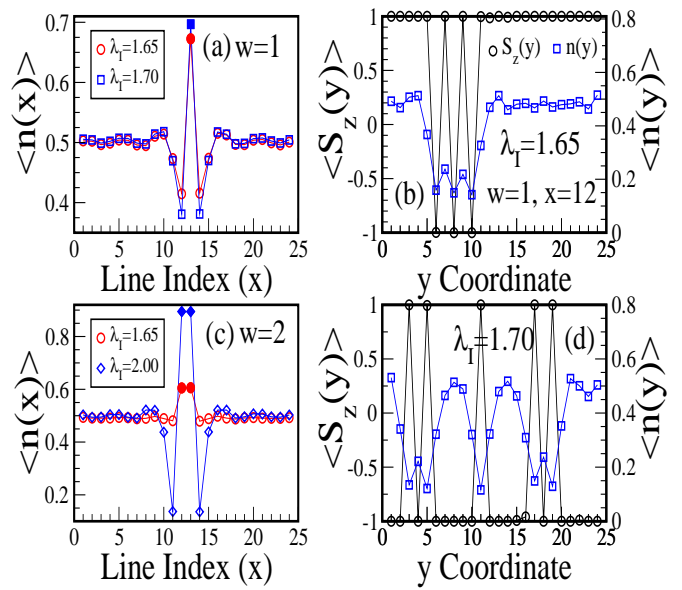}}
\caption{(Color online) Line averaged electron density $\langle n(x) \rangle $ for (a) $w=1$, 
(c) $w=2$. The $z$ component of the t$_{2g}$ spins $\langle S_z(y) \rangle$ and $\langle n(y) \rangle$ 
for each site of the FM interfacial line $x=12$ for $w=1$ using (b) $\lambda_I=1.65$ and (d) $\lambda_I=1.70$. 
Legends in (b) and (d) are the same. Open and closed symbols in (a) and (c) are from lines in the FM and 
the AFI layers, respectively. 
}
\end{figure}

$\langle S_I(\textbf{0}) \rangle$ for different $\lambda_I$ values also varies non-monotonically with 
increasing width of the AFI layer except for $\lambda_I$ = 2.0. The induced magnetic moment for $w>2$ 
decreases more rapidly for larger electron-phonon coupling $\lambda_I$. The AFI layer recovers 
the AF, charge ordered state at a smaller width $w$ for larger $\lambda_I$. 
This is why it is possible to magnetize only 2 lines of the AFI layer for $\lambda_I=1.8$ while for 
$\lambda_I=2.0$ the induced magnetic moment remains very small in the AFI layer irrespective 
of its width. 

In order to understand the non-monotonic behavior we specifically choose $w=1$ and $w=9$ for which 
$\langle S_I(\textbf{0}) \rangle$ is small. To start with we analyze the magnetization profile across 
the interface by calculating average magnetization for each line of the superlattice $\langle S_z(x) \rangle$ 
for transverse coordinate $x$. $\langle S_z(x) \rangle$, in Fig.2(c), decreases for $x$ = 11-15 for $w=9$, 
$i.e.$ in the center lines of the AFI layer, 
which implies that the induced ferromagnetic moment in the AFI layer is confined to the near vicinity 
of the interface. The relation between the induced magnetization and the line-averaged electron density 
$\langle n(x) \rangle$ becomes evident in Fig.2(d). $\langle n(x) \rangle$ for the interfacial line on 
the FM side, named as FM interfacial line, decreases while the AFI interfacial line increases from the 
initial electron density 0.5 to $\sim$0.65. The induced ferromagnetic moment for the lines $x$ = 9 and 
17 is therefore due to the enhanced electron density and the spin bias from the ferromagnetic metal. 
In fact, for the parameters $J=0.1$, $n=0.65$, and $\lambda_I=1.65$ the groundstate of the bulk system 
is a FM. The magnetization in the line $x$ = 10 (16) is induced by the fully magnetized line $x$ = 9 (17).
The interfacial lines of the AFI layer are also magnetized for other $w$ values, except for $w=1$
which is discussed later. 
The spin bias from the ferromagnetic metal is important for the induction of a ferromagnetic moment in the 
AFI interfacial lines. The induced magnetization in the AFI layer is very small irrespective of the AFI layer 
width where the FM interfacial lines are magnetically disordered\cite{kp-sl-prb}.

The direction of electron transfer is from the FM to the AFI layer as anticipated earlier. Sufficiently far 
away from the interface the average electron density must return to the initial electron density $n=0.5$, 
which however is not fully accomplished for $w=9$ and $\alpha=0$. But with the additional LRC interaction 
$\langle n(x) \rangle$ indeed gradually returns to the initial electron density (see Fig.2(d)). For 
$\alpha=0.1$ the average electron densities are clearly higher (lower) in the FM (AFI) layers as compared to 
$\alpha=0$. In the FM/AFI superlattices, where the constituent FM and AFI manganites have the same initial 
electron density, the LRC interaction reduces the critical width, beyond which $\langle S_I({\textbf 0}) \rangle$ 
starts to decrease\cite{kp-sl-prb}. Remarkably, $\langle n(x) \rangle$ at the FM interfacial lines is largely 
unaltered by the LRC interactions.

The line averaged $\langle S_z(x) \rangle$ for $w=1$ is also shown in Fig.2(c). But in contrast to 
the AFI layer width $w=9$, $\langle S_z(x) \rangle$ in the FM interfacial line decreases for $w=1$. 
The difference results from the decrease in the electron density for $w=1$ in the FM interfacial line as 
shown in Fig.3(a). The spin pattern in the interfacial line decomposes into FM and G-type AF regions. 
This is shown in Fig.3(b) which displays the averaged $z$ components of the t$_{2g}$ spins 
$\langle S_z(y) \rangle$ for each site of the FM interfacial line $x=12$ for one selected 'sample'. 

The averaged electron density at the FM interfacial line is smaller for $\lambda_I=1.70$ as compared 
to $\lambda_I=1.65$ as shown in Fig.3(a). For this reason the G-type AF regions in the FM interfacial 
line are more pronounced for $\lambda_I=1.70$ (see Fig.3(d)). The electron densities and the $z$ components 
of the t$_{2g}$ spins at each site in the FM/AFI superlattice are shown in Fig.4 for $w=1$. The magnetic and 
the electronic profile of both FM interfacial lines are similar to each other on both sides of the AFI 
line. The magnetic profile of the AFI line is tied to the profile of the FM interfacial lines while the 
electron density of the sites in the AFI layer is enhanced to $\sim$0.7. 

With increasingly larger values of $\lambda_I$ the magnetization of the FM interfacial line decreases due to the 
enhanced G-type correlations for $w=1$. This establishes the crucial relation between the magnetization at the 
FM interfacial line and the induced magnetic moment in the AFI layer. This is in general true for any width $w$. 
So the non-monotonic behavior of $\langle S_I(\textbf{0}) \rangle$ in Fig.2(a) implies that the FM interfacial line 
remains ferromagnetic for $w=2$. For $w=2$ the decrease in the electron density in the FM interfacial 
line (see Fig.3(c), $\lambda_I=1.65$) is very small as compared to $w=1$, and these lines therefore remain 
ferromagnetic. For $\lambda_I=1.80$ the electron density in the FM interfacial line decreases considerably
except for $w=2$. The electron density profile for $w=2$ and $\lambda_I=1.80$ resembles the profile of
$w=2$ and $\lambda_I=1.65$ shown in Fig.3(c). This implies that the charge transfer across the interface from
the FM to the AFI layer also varies non-monotonically similar to the induced magnetic moment
in the AFI layer shown in Fig.2(a). The competition between FM and G-type AF spin patterns at the interface is
controlled by the double-exchange energy gain due to the induced magnetic moment in the AFI layer. The
magnetization of the FM interfacial layer remains ferromagnetic for $w=2$, i.e. the large induced
magnetic moment in the AFI layer removes the MDL.

\begin{figure} [t]
\centerline{
\includegraphics[width=8.3cm,height=3.9cm,clip=true]{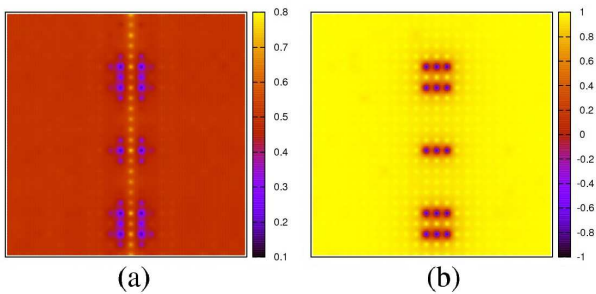}}
\caption{(Color online) (a) The electron densities and (b) the $z$ components of the $t_{2g}$ spins 
for each site on a 24$\times$24 superlattice at $T=0.01$ with the AFI layer width $w=1$ 
($\lambda_M=1.0$ and $\lambda_I=1.7$).
}
\end{figure}

At large electron-phonon couplings $\lambda_I\gtrsim 2.0$ electrons are site-localized due to strong lattice 
distortions; this decreases the double-exchange energy gain from induced ferromagnetic moments which are 
hence absent in the AFI layer (see Fig.2(a)) irrespective of the AFI 
layer width. For this reason G-type spin patterns are more 
prominent at the FM interfacial lines and $\langle n(x) \rangle$ in the FM interfacial line decreases considerably 
(see e.g. Fig.3(c) for $w=2$ and $\lambda_I=2.0$). These results suggest that also in FM/NMI superlattices 
local AF correlations may emerge in the FM interfacial line and the magnetization at the interface is wiped out 
originating in an MDL due to the decrease in the electron density at the interface\cite{sun-NdGaO3-mdl,bibes-lcmo-mdl}.

In the type I FM/AFI superlattices the spins in the FM leads are aligned in the same direction due to the 
periodic boundary conditions; this set up mimics the experimental situation in which the FM layers of 
the superlattice are aligned by an external magnetic field. Specifically we have designed the type II superlattice 
where two AFI layers instead of one are considered as shown in Fig.1(b) to represent more closely the experimental 
setup. The magnetizations in the left and the right 
FM layers are aligned parallel while the middle FM layer is free to choose its spin direction. A small external 
magnetic field $h$ is applied to align all the FM layers in the same direction. Fig.5(a) shows 
the line averaged $\langle n(x) \rangle$ vs. line index $x$ for $w=1$, $\lambda_I=1.8$, and $h=0.002$. In Fig.5(b), 
we plot the averaged $<S_{zI}>$ in the AFI layers along with averaged $z$ component of the $t_{2g}$ spins in the 
FM interfacial lines $<S_{z}(IL)>$ for the same magnetic field. The magnetization of the FM interfacial line follows 
a non-monotonic behavior similar to the induced magnetic moments in the AFI layer. The dc limit of the 
longitudinal conductivity $\sigma_{dc}$, also displayed in Fig.5(b), as obtained from the Kubo-Greenwood 
formula\cite{mahan-book,cond-ref}, follows the same trend with increasing AFI layer width $w$. It is the 
combination of the induced magnetic moment in the AFI layer and the magnetization of the FM interfacial lines 
which enhances the conductivity. 

\begin{figure} [t]
\centerline{
\includegraphics[width=8.75cm,height=3.5cm,clip=true]{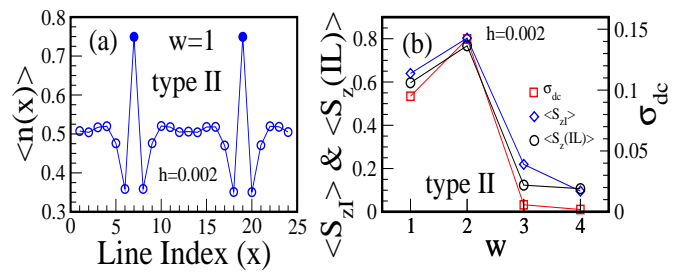}}
\caption{(Color online) Type II FM/AFI superlattice ($\lambda_M=1.0$ and $\lambda_I=1.8$) : (a) Line 
averaged $z$ component of the t$_{2g}$ spins $\langle S_z(x) \rangle$ for $w=1$. 
(b) $\langle S_{zI} \rangle$, $\langle S_{z}(IL)\rangle$ (see text) and the dc conductivity 
(in units of ${\pi  e^2 }/{\hbar a}$ where a is the lattice spacing, see Ref.\citenum{cond-ref}) at 
$T=0.01$ for different AFI layer widths $w$. An external magnetic field $h=0.002$ is applied to align 
the FM layers.
}
\end{figure}

In the type II set up the TMR may be calculated by fixing the spins of the middle FM layer in the direction opposite 
to that of the left and right FM layers. However in the limit $J_H \rightarrow \infty$ adopted here where the spins of 
the mobile e$_g$ electrons are perfectly aligned along the local $t_{2g}$ spin direction the dc conductivity 
(resistivity) for this set up is zero (infinity). For this reason a quantitative calculation of TMR for different
widths $w$ of the AFI layer is not presented here. In the experiments the resistivity is large but finite in the
antiparallel configuration of the FM layers. The increase of the conductivity in the parallel configuration of
the FM layers, shown in Fig.5(b) will necessarily enhance the TMR.

In conclusion, our 2D model calculations provide a framework to explain the origin of the MDL at the FM 
interface in FM/Insulator superlattices. The magnetization of the interfacial lines of the FM layers is determined by 
the amount of electron transfer from the FM interfacial lines to the AFI layer. The decrease in the 
magnetization of the FM interface, when joined with a NMI oxide is due to the decrease in the electron density 
at the interfacial lines as a result of the charge transfer across the interface. The amount of transferred charge 
is limited in a scenario for which instead AFI layers are sandwiched between FM layers, since inducing ferromagnetic 
moment in the AFI layer requires to control the charge transfer. But even in such a FM/AFI superlattices, the MDL is absent 
only for a specific range of AFI layer widths, because the induced magnetic moment in the AFI layer varies 
non-monotonically with the AFI layer width\cite{mathur-lsmo-pcmo}. The absence of the MDL in FM/AFI superlattices
enhances the TMR. The MDL at the interface in an FM/NMI junction may be minimized by the insertion of an intervening AFI
layer. In such
a setup, the width of the AFI layer has to be chosen such that the AFI layer is maximally polarized along the direction
of the magnetization in the FM layers due to charge transfer. Indeed the TMR is significantly enhanced in the
engineered FM/NMI MTJs with an intervening AFI layer\cite{fm-afi-nmi-ref,fm-afi-nmi-ref1}. The role of the MDL
for different widths of the intervening AFI layer in these engineered MTJs deserves further investigation. We
leave this as the subject for future work.

This work was supported by the DFG through TRR80. 



\end{document}